\documentclass[twocolumn,english,superscriptaddress,citeautoscript,showpacs,preprintnumbers,amsmath,amssymb,prb,floatfix,footinbib]{revtex4}
\usepackage[scaled=2]{helvet}
\usepackage[T1]{fontenc}
\usepackage[latin9]{inputenc}
\usepackage{textcomp}
\usepackage{amsmath}
\usepackage{amssymb}
\usepackage{graphicx}
\usepackage{subscript}

\makeatletter

\providecommand{\tabularnewline}{\\}

\@ifundefined{textcolor}{}
{%
 \definecolor{BLACK}{gray}{0}
 \definecolor{WHITE}{gray}{1}
 \definecolor{RED}{rgb}{1,0,0}
 \definecolor{GREEN}{rgb}{0,1,0}
 \definecolor{BLUE}{rgb}{0,0,1}
 \definecolor{CYAN}{cmyk}{1,0,0,0}
 \definecolor{MAGENTA}{cmyk}{0,1,0,0}
 \definecolor{YELLOW}{cmyk}{0,0,1,0}
}

\usepackage[scaled=2]{helvet}\usepackage{amstext}

\makeatletter%%%%%%%%%%%%%%%%%%%%%%%%%%%%%% Textclass specific LaTeX commands.
\@ifundefined{textcolor}{}{\definecolor{BLACK}{gray}{0}\definecolor{WHITE}{gray}{1}\definecolor{RED}{rgb}{1,0,0}\definecolor{GREEN}{rgb}{0,1,0}\definecolor{BLUE}{rgb}{0,0,1}\definecolor{CYAN}{cmyk}{1,0,0,0}\definecolor{MAGENTA}{cmyk}{0,1,0,0}\definecolor{YELLOW}{cmyk}{0,0,1,0}}

\usepackage{dcolumn}% Align table columns on decimal point
\usepackage{bm}% bold math
\usepackage{times}% font that prx use
\usepackage{hyperref}
\hypersetup{colorlinks=true,linkcolor=red,citecolor=blue}
\makeatother

\usepackage{babel}

\makeatother

\usepackage{babel}
\begin{document}

\title{Magnetic structure of superconducting Eu(Fe\textsubscript{0.82}Co\textsubscript{0.18})\textsubscript{2}As\textsubscript{2}
as revealed by single-crystal neutron diffraction}

\author{W. T. Jin}

\email{w.jin@fz-juelich.de}

\affiliation{J\"{u}lich Centre for Neutron Science JCNS and Peter Gr\"{u}nberg Institut
PGI, JARA-FIT, Forschungszentrum J\"{u}lich GmbH, D-52425 J\"{u}lich, Germany}

\affiliation{J\"{u}lich Centre for Neutron Science JCNS, Forschungszentrum J\"{u}lich
GmbH, Outstation at MLZ, Lichtenbergstra\ss{}e 1, D-85747 Garching, Germany}

\author{S. Nandi}

\affiliation{J\"{u}lich Centre for Neutron Science JCNS and Peter Gr\"{u}nberg Institut
PGI, JARA-FIT, Forschungszentrum J\"{u}lich GmbH, D-52425 J\"{u}lich, Germany}

\affiliation{J\"{u}lich Centre for Neutron Science JCNS, Forschungszentrum J\"{u}lich
GmbH, Outstation at MLZ, Lichtenbergstra\ss{}e 1, D-85747 Garching, Germany}

\author{Y. Xiao}

\affiliation{J\"{u}lich Centre for Neutron Science JCNS and Peter Gr\"{u}nberg Institut
PGI, JARA-FIT, Forschungszentrum J\"{u}lich GmbH, D-52425 J\"{u}lich, Germany}

\author{Y. Su}

\affiliation{J\"{u}lich Centre for Neutron Science JCNS, Forschungszentrum J\"{u}lich
GmbH, Outstation at MLZ, Lichtenbergstra\ss{}e 1, D-85747 Garching, Germany}

\author{O. Zaharko}

\affiliation{Laboratory for Neutron Scattering, Paul Scherrer Institut, CH-5232
Villigen PSI, Switzerland}

\author{Z. Guguchia}

\affiliation{Physik-Institut der Universit\"{a}t Z\"{u}rich, Winterthurerstrasse 190,
CH-8057 Z\"{u}rich, Switzerland}

\author{Z. Bukowski }

\affiliation{Institute of Low Temperature and Structure Research, Polish Academy
of Sciences, 50-422 Wroclaw, Poland}

\affiliation{Laboratory for Solid State Physics, ETH Z\"{u}rich, CH-8093 Z\"{u}rich, Switzerland}

\author{S. Price}

\affiliation{J\"{u}lich Centre for Neutron Science JCNS and Peter Gr\"{u}nberg Institut
PGI, JARA-FIT, Forschungszentrum J\"{u}lich GmbH, D-52425 J\"{u}lich, Germany}

\author{W. H. Jiao}

\affiliation{Department of Physics, Zhejiang University, Hangzhou 310027, China}

\author{G. H. Cao}

\affiliation{Department of Physics, Zhejiang University, Hangzhou 310027, China}

\author{Th. Br\"{u}ckel}

\affiliation{J\"{u}lich Centre for Neutron Science JCNS and Peter Gr\"{u}nberg Institut
PGI, JARA-FIT, Forschungszentrum J\"{u}lich GmbH, D-52425 J\"{u}lich, Germany}

\affiliation{J\"{u}lich Centre for Neutron Science JCNS, Forschungszentrum J\"{u}lich
GmbH, Outstation at MLZ, Lichtenbergstra\ss{}e 1, D-85747 Garching, Germany}

\begin{abstract}
The magnetic structure of superconducting Eu(Fe\textsubscript{0.82}Co\textsubscript{0.18})\textsubscript{2}As\textsubscript{2} is unambiguously determined by single crystal neutron diffraction. A long-range ferromagnetic order of the Eu\textsuperscript{2+} moments along the \textit{c}-direction is revealed below the magnetic phase transition temperature $T_{C}$ = 17 K. In addition, the antiferromagnetism of the Fe\textsuperscript{2+} moments still survives and the tetragonal-to-orthorhombic structural phase transition is also observed, although the transition temperatures of the Fe-spin density wave (SDW) order and the structural phase transition are significantly suppressed to $T_{N}$ = 70 K and $T_{S}$ = 90 K, respectively, compared to the parent compound EuFe\textsubscript{2}As\textsubscript{2}. We present the microscopic evidences for the coexistence of the Eu-ferromagnetism (FM) and the Fe-SDW in the superconducting crystal. The superconductivity (SC) competes with the Fe-SDW in Eu(Fe\textsubscript{0.82}Co\textsubscript{0.18})\textsubscript{2}As\textsubscript{2}. Moreover, the comparison between Eu(Fe$_{1-x}$Co$_{x}$)\textsubscript{2}As\textsubscript{2} and Ba(Fe$_{1-x}$Co$_{x}$)\textsubscript{2}As\textsubscript{2} indicates a considerable influence of the rare-earth element Eu on the magnetism of the Fe sublattice.

\end{abstract}


\maketitle

\section{Introduction}

Since the discovery of Fe-pnictide superconductors in 2008,\cite{kamihara_08} a great deal of attention has been given to the investigation and understanding of the interplay between magnetism and superconductivity in these new materials.\cite{johnston_10,lumsden_10,dai_12} The parent compounds of Fe-pnictides undergo a structural phase transition from tetragonal to orthorhombic, accompanied \cite{huang_08} or followed \cite{Cruz_08} by an antiferromagnetic spin-density-wave (SDW) transition. Superconductivity can be induced by doping the parent compounds with charge carriers, \cite{rotter_08,sefat_08} or by applying the hydrostatic or internal chemical pressure. \cite{torikachvili_08,ren_EuP_09} Meanwhile, both magnetic order and structural distortion are suppressed. Although this is a general tendency common for different compounds, the structural and physical behavior near the phase boundary between the antiferromagnetic (AFM) and the superconducting (SC) phases is complex and material specific. For some compounds such as CeFeAsO$_{1-x}$F$_{x}$, the AFM and the SC phases seem mutually exclusive.\cite{zhao_08} However, in some other materials like Ba(Fe$_{1-x}$Co$_{x}$)\textsubscript{2}As\textsubscript{2}, the AFM and the SC phases coexist and compete with each other.\cite{pratt_09,Christianson_09} The proximity between SC and AFM resembles that in cuprates and heavy fermion systems, suggesting that the SC in the Fe-pnictide is also unconventional and that magnetism might play a role in the underlying mechanism. 

EuFe\textsubscript{2}As\textsubscript{2} is a unique member of the ternary iron arsenide $A$Fe\textsubscript{2}As\textsubscript{2} (\textquotedblleft{}122\textquotedblright{}, $A$\,=\,Alkaline earth or rare-earth) family, since the $A$ site is occupied by an \textsl{S}-state
(orbital moment $L$ = 0) rare-earth Eu\textsuperscript{2+} possessing a 4\textsl{f}\textsuperscript{7} electronic configuration with an electron spin \textsl{S} = 7/2, corresponding to a theoretical effective magnetic moment of 7.94 $\mu_{B}$.\cite{marchand_78} Interestingly,
it was found that both magnetic sublattices in the unit cell, Fe and Eu layers, order antiferromagnetically below 190 K and 19 K, respectively.\cite{raffius_93,ren_EuFeAs08,jeevan_EuFeAs08} Further studies using magnetic resonant x-ray scattering \cite{Herrero-Martin_09} and neutron diffraction \cite{Xiao_09} confirmed that the AFM of Eu\textsuperscript{2+} spins is of A-type, i.e., ferromagnetic layers with the Eu\textsuperscript{2+} moments aligned along the \textit{a}-axis order antiferromagnetically along the \textsl{c}-direction. The Fe\textsuperscript{2+} moments were revealed to order antiferromagnetically along the orthorhombic\textit{ a}-axis.

Similar to other Fe-pnictides, superconductivity can be achieved in the EuFe\textsubscript{2}As\textsubscript{2} family by chemical substitution at different sites \cite{ren_EuP_09,jeevan_EuK08,jiang_09,jiao_EPL09} or by application of external pressure.\cite{Miclea_09} It is well established that in the doped-EuFe\textsubscript{2}As\textsubscript{2} system, similar to other doped-122 families, the SDW order of the Fe\textsuperscript{2+} moments gets gradually suppressed with an increase of the doping level, \cite{jeevan_EuK08,jeevan_EuP11,Ren_EuNi09,Blachowski_11,zhang_EuLa12,jiao_EuRu12} in favor of the occurrence of SC. The Fe-SDW order is suppressed \cite{jeevan_EuK08,jeevan_EuP11,Ren_EuNi09} or coexists with the SC within a certain doping regime \cite{Blachowski_11,zhang_EuLa12,jiao_EuRu12} depending on the dopants. Moreover, the Fe-SDW order and the orthorhombic distortion exhibit very weak coupling with the magnetic order of Eu\textsuperscript{2+} spins based on the result from the undoped EuFe\textsubscript{2}As\textsubscript{2} parent compound.\cite{Xiao_09} However, the evolution of the magnetic ordering of Eu\textsuperscript{2+} spins with increasing doping level and its interplay with the SC is still not completely clarified. For EuFe\textsubscript{2}(As$_{1-x}$P$_{x}$)\textsubscript{2} with isovalent P doping on the As site, it is generally recognized that the magnetic moments of Eu\textsuperscript{2+} evolve from the A-type AFM order at low doping level, to the ferromagnetic order at high doping level, although the magnetic structure of Eu\textsuperscript{2+} spins in the superconducting region of the phase diagram is quite controversial.\cite{jeevan_EuP11,cao_11,zapf_11} Recently, by combination of magnetization, specific heat and magnetic resonant x-ray scattering measurements, we conclude that in an EuFe\textsubscript{2}(As$_{1-x}$P$_{x}$)\textsubscript{2} single crystal with $x$ = 0.15, the Eu\textsuperscript{2+} magnetic moments order ferromagnetically primarily along the \textsl{c}-axis and the ferromagnetism (FM) coexists with bulk SC. \cite{nandi_13} However, for Eu(Fe$_{1-x}$Co$_{x}$)\textsubscript{2}As\textsubscript{2} , so far there is still no clear picture regarding how the magnetic ordering of the Eu\textsuperscript{2+} spins develops with increasing Co concentration and it is even more controversial compared with the P-doped case. For instance, there exist several different proposals for the magnetic ordering of Eu\textsuperscript{2+} around 10\% Co concentration including the in-plane helical structure \cite{jiang_09} and the canted structure with a ferromagnetic component in the \textit{a-b} plane \cite{guguchia_11} or along the \textit{c} direction. \cite{nowik_11} 

To our knowledge, for Eu(Fe$_{1-x}$Co$_{x}$)\textsubscript{2}As\textsubscript{2}, direct microscopic determination of the magnetic structure under zero magnetic field is still lacking. The neutron diffraction technique stands out due to its ability to probe the bulk and the high accuracy in determining both nuclear and magnetic structures. However, neutron experiments on Eu-containing materials are difficult and challenging due to the large neutron absorption cross section of Eu. Nevertheless, by significant reduction of the absorption effect using short-wavelength neutrons, such experiments prove feasible for a crystal of good quality. Here we present the results of our neutron diffraction measurements on a high-quality superconducting Eu(Fe\textsubscript{0.82}Co\textsubscript{0.18})\textsubscript{2}As\textsubscript{2} single crystal, which indicate unambiguously that the Eu\textsuperscript{2+} moments are long-range ferromagnetically ordered, oriented purely along the\textsl{ c}-axis. Surprisingly, very weak magnetic reflections arising from the remaining antiferromagnetism of the Fe moments are also observed. Therefore, it is revealed that both the Eu-FM and the Fe-SDW coexist with the SC in Eu(Fe\textsubscript{0.82}Co\textsubscript{0.18})\textsubscript{2}As\textsubscript{2}.

\section{Experimental Details}

Single crystals of Eu(Fe\textsubscript{0.82}Co\textsubscript{0.18})\textsubscript{2}As\textsubscript{2} were grown out of Sn flux. \cite{Guguchia_11_NMR} The chemical composition of this batch was determined by wavelength dispersive spectroscopy (WDS). X-ray Laue diffraction confirmed the high quality of the crystals with the \textsl{c}-axis perpendicular to their surfaces. A 100 mg platelet-like single crystal with dimensions \textasciitilde{}5$\times$5$\times$1.5 mm\textsuperscript{3} was selected for neutron diffraction measurements, which were carried out on the thermal-neutron four-circle diffractometer TriCS \cite{schefer_00} at the Swiss Spallation Source (SINQ). The single-crystal sample was mounted on an aluminum sample holder with a small amount of GE varnish. The sample holder was then mounted inside a small Al-can filled with Helium exchange gas, allowing it to reach a base temperature of 4.5 K. A Ge (3 1 1) monochromator was chosen to produce a monochromatic neutron beam with the wavelength of 1.178 $\buildrel_\circ \over {\mathrm{A}}$, for which the neutron absorption cross-section of Eu is 2965 barn. The diffracted neutron beam was collected with a \textsuperscript{3}He single detector. In order to determine the nuclear and magnetic structure, the integrated intensities of 348 reflections at 4.5 K and 330 reflections at 25 K (above the magnetic ordering temperature of the Eu\textsuperscript{2+} moments) were collected for refinements without a collimator in front of the detector. For the measurements of the weak reflections, the temperature dependencies, and the Q-scans, a collimation of 40\'{ } in front of the detector was installed to suppress background. The obtained reflection sets at both temperatures were normalized to the monitor and corrected for the Lorentz factor. DATAP program was used for the absorption correction by considering the size and shape of the crystal.\cite{coppens_65} Refinement of both nuclear and magnetic structures was carried out using the FULLPROF program suit.\cite{Rodriguez_93} For macroscopic characterizations, a small plate-like crystal of 9.4 mg from the same batch was chosen. The resistivity and magnetization were measured using a Quantum Design physical property measurement system (PPMS) and a Quantum Design magnetic property measurement system (MPMS), respectively.

\section{Results and discussion}

\begin{figure}
\centering{}\includegraphics{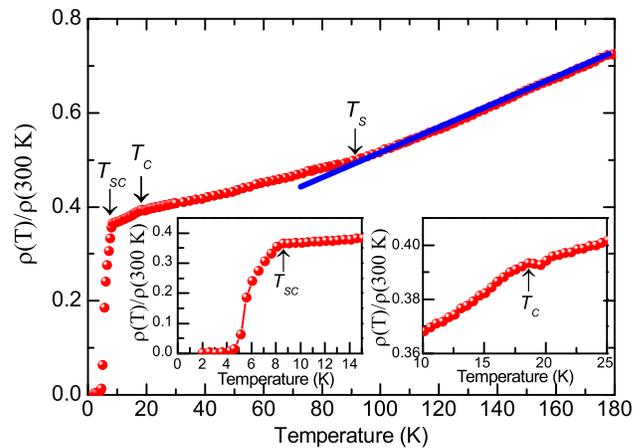}

\caption{The temperature dependence of the in-plane electrical resistivity of Eu(Fe\textsubscript{0.82}Co\textsubscript{0.18})\textsubscript{2}As\textsubscript{2}. The blue solid line is the linear fitting of the high-$T$ resistivity and $T_{S}$ marks the temperature where experiment starts to deviate from the linear behavior. $T_{C}$ and $T_{SC}$ denote the magnetic ordering temperature of the Eu\textsuperscript{2+} moments and the superconducting transition temperature, respectively. The two insets give an enlarged illustration of the \textit{R-T} curve around $T_{SC}$ and $T_{C}$, respectively.} 
\end{figure}

The temperature dependence of the in-plane electrical resistivity of the Eu(Fe\textsubscript{0.82}Co\textsubscript{0.18})\textsubscript{2}As\textsubscript{2} single crystal is shown in Figure 1. The resistivity descends smoothly with decreasing temperature, reflecting
its metallic behavior. Above 90 K, the resistivity exhibits a linear temperature dependence. The slope of \textsl{R-T} curve changes below 90 K and a pronounced kink emerges there, probably corresponding to the change in Fermi surface nesting features due to the structural
distortion, as in all other Fe-pnictides. We denote 90 K as $T_{S}$ since it coincides with the structural phase transition temperature determined by neutron measurements. Around 17 K (denoted as $T_{C}$), another kink appears, which is clearer as shown in the right inset
of Fig. 1, corresponding to the magnetic ordering of the Eu\textsuperscript{2+} spins. This is similar to the reentrant resistivity reported in Ref. \onlinecite{jiang_09} and \onlinecite{he_10}, but here it is due to the ferromagnetic ordering of Eu\textsuperscript{2+} moments as evidenced by our neutron data, which will be presented below. Below 8 K (denoted as $T_{SC}$), the resistivity drops sharply and finally a zero-resistivity superconducting state (less than $10^{-8}$ $\Omega\cdot m^{-1}$) is achieved below 4 K, as illustrated in the left inset of Fig. 1.The SC transition in Co-doped EuFe\textsubscript{2}As\textsubscript{2} is usually more susceptible to the adverse effect of the Eu-magnetic ordering and zero resistivity is only realized within a quite narrow Co-concentration window around 20\%.\cite{Blachowski_11} The \textsl{R-T} behavior here is very similar to that reported by Tran \textsl{et al.} on a single crystal with a similar composition Eu(Fe\textsubscript{0.81}Co\textsubscript{0.19})\textsubscript{2}As\textsubscript{2}, \cite{tran_12} in which also multiple phase transitions with comparable transition temperatures were found. 

\begin{figure}

\centering{}\includegraphics{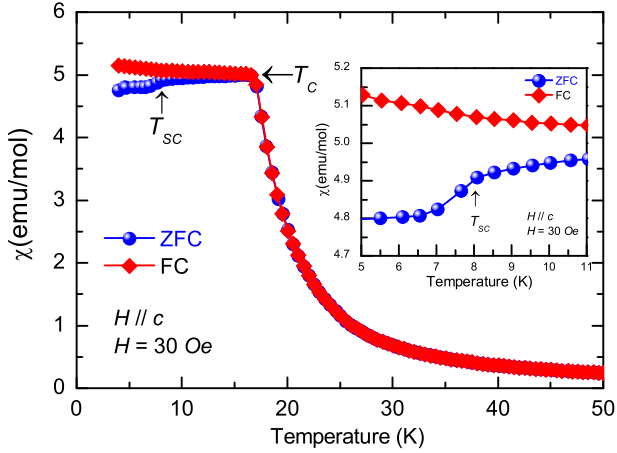}\caption{The temperature dependence of the magnetic susceptibility of Eu(Fe\textsubscript{0.82}Co\textsubscript{0.18})\textsubscript{2}As\textsubscript{2 }measured in an applied field of 30 Oe along the \textit{c}-direction in ZFC and FC process, respectively. The inset is the enlarged illustration of the ZFC and FC curves around $T_{SC}$.}

\end{figure}

Figure 2 shows the temperature dependence of the magnetic susceptibility ($\chi$) for the Eu(Fe\textsubscript{0.82}Co\textsubscript{0.18})\textsubscript{2}As\textsubscript{2} single crystal below 50 K under an applied field of 30 Oe along the \textsl{c}-direction. A bifurcation between zero-field-cooling (ZFC) and field-cooling (FC) curves develops below 17 K, indicating the emergence of a FM ordered state. Upon further cooling, a sudden drop around 8 K occurs for the ZFC curve, which results from the diamagnetic response of the SC transition. However, negative susceptibility is not achieved due to the small superconducting volume and the dominance of the Eu-FM over the SC. No obvious Meissner effect is detected below 8 K by the FC measurement, similar to the case in Eu(Fe$_{0.88}$Ir$_{0.12}$)\textsubscript{2}As\textsubscript{2}, ascribing to the strong internal field  induced by the Eu-FM. \cite{Jiao_13} The anomaly in susceptibility due to the Fe-SDW transition around $T_{S}$ is hardly observed (not shown here) due to the small size of the Fe moments, even after subtracting the Curie-Weiss contribution of the Eu\textsuperscript{2+} moments. 

Hinted by the kink around 90 K in the \textsl{R-T} curve in Fig. 1, it seems that Eu(Fe\textsubscript{0.82}Co\textsubscript{0.18})\textsubscript{2}As\textsubscript{2} also undergoes a structural phase transition from tetragonal (space group \textit{I4/mmm}) to orthorhombic (\textit{Fmmm}) while cooling, similar to the parent compound EuFe\textsubscript{2}As\textsubscript{2}. Here we present evidence for the occurrence of the structural phase transition in this compound from our neutron data. Figure 3(a) illustrates the rocking curve scans ($\omega$ scans) of the (-4 0 0)/(0 -4 0) reflection at 120 K, 85 K and 5 K, respectively. This corresponds to the (-2 2 0)\textsubscript{$T$} reflection in tetragonal notation. For convenience, we use the orthorhombic notation with the shortest axis defined as \textit{b }throughout this paper. The mosaic width of less than 0.3\textdegree{} confirms the good quality of the crystal. Upon cooling, the splitting of this peak into two distinct peaks could not be resolved due to limited instrumental resolution and the intrinsically small orthorhombic distortion. However, the structural phase transition indeed occurs based on the temperature dependencies of both the integrated intensity and the full width at half maximum (FWHM) of the (-4 0 0)/(0 -4 0) peak, as shown in Fig. 3(b) and Fig. 3(c), respectively. The rapid
increase of the intensity and broadening of the width of (-4 0 0)/(0 -4 0) peak below 90 K reflect the structural change from the tetragonal phase to the orthorhombic phase. This is similar to the observation in Ba(Fe$_{1-x}$Co$_{x}$)\textsubscript{2}As\textsubscript{2} reported by Lester \textit{et al}.,\cite{lester_09} where no orthorhombic splitting was resolved but an obvious kink at $T_{S}$ was observed in the temperature dependence of the integrated intensity of the (2 2 0)\textsubscript{T} peak. The structural transition temperature $T_{S}$ determined here is in good agreement with that shown in the resistivity measurement, being around 90 K. 

\begin{figure}

\centering{}\includegraphics{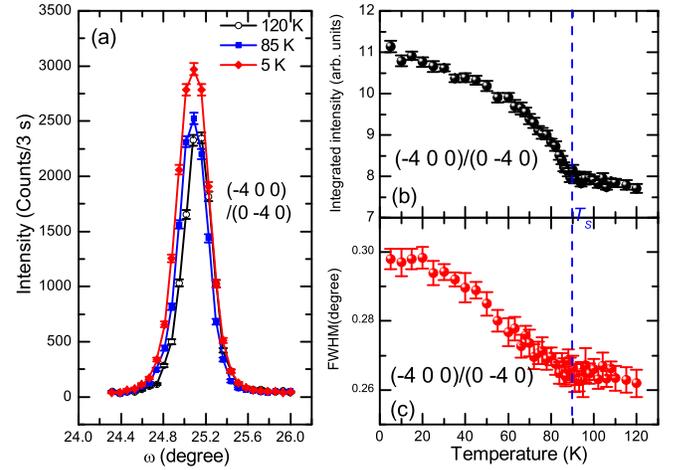}

\caption{(a) The rocking curve scans ($\omega$ scans) of (-4 0 0)/(0 -4 0) reflection at 120 K, 85 K and 5 K, respectively. The temperature dependence of the integrated intensity (b) and the peak width (FWHM) (c) of (-4 0 0)/(0 -4 0) peak both show a kink around $T_{S}$ = 90 K (marked by the blue vertical dashed line) corresponding to the tetragonal-orthorhombic structural phase transition.}
\end{figure}

Fig. 4(a) and Fig. 4(b) show the comparison of the (-2 0 0) and (1 1 1) peaks between 4.5 K and 25 K. Both are very weak nuclear reflections at 25 K which is above the magnetic ordering temperature of the Eu\textsuperscript{2+} moments. Upon cooling, the magnetic contribution from the Eu\textsuperscript{2+} magnetic ordering develops on top of the nuclear part. At 4.5 K, both the (-2 0 0) and (1 1 1) peak become extremely strong, indicating a large ferromagnetic component from the Eu\textsuperscript{2+} moments perpendicular to the scattering vectors. The obvious difference of the (-2 0 2) peak between 4.5 K and 25 K (Fig. 4(c)) also suggests a ferromagnetic contribution, although its nuclear part is quite strong. On the other hand, the (0 0 2), (0 0 4) and (0 0 8) peaks show no discernible difference between the two temperatures(Fig. 4(d-f)), suggesting a very small or even absent in-plane ferromagnetic component of the Eu\textsuperscript{2+} magnetic moment. Hence we can conclude that the ferromagnetic component of the Eu\textsuperscript{2+} moments lies in \textit{c} direction within our experimental accuracy. Fig. 4(g) illustrates the temperature dependencies of the integrated intensities of the (-2 0 0), (1 1 1) and (-2 0 2) peaks. The net increase on top of the nuclear contribution upon cooling represents the order parameter of the ferromagnetic transition, which can be well fitted by a power law $I-I_{0}\varpropto(1-T/T_{C})^{2\beta}$, yielding the transition temperature $T_{C}$ = 16.9(2) K and the exponent $\beta$ = 0.35(2). $T_{C}$ determined here is also in good agreement with the results from resistivity and magnetization measurements. The exponent $\beta$ is similar to that of the incommensurate antiferromagnetic ordering of the Eu\textsuperscript{2+} moments in EuRh\textsubscript{2}As\textsubscript{2} ($\beta$ = 0.32 $\pm$ 0.02).\cite{nandi_09} Both exponents are close to the critical exponent of the three-dimensional classical Heisenberg model ($\beta$ = 0.36), typical for the rare-earth elements in intermetallic compounds.\cite{Brueckel_01} However, here the power law refinement holds over an unusual wide temperature range (down to 7 K when the order parameter tends to saturate), well outside the usual critical region. In addition, no evident anomaly is observed around the superconducting temperature $T_{SC}$ (\textasciitilde{}8 K) in the temperature dependence of the Eu-FM order parameter as shown in Fig. 4(g). However, considering the small superconducting volume in Eu(Fe\textsubscript{0.82}Co\textsubscript{0.18})\textsubscript{2}As\textsubscript{2} as shown in the magnetic susceptibility measurement, it is difficult to conclude about the interplay between superconductivity and Eu-FM.

\begin{figure}
\centering{}\includegraphics{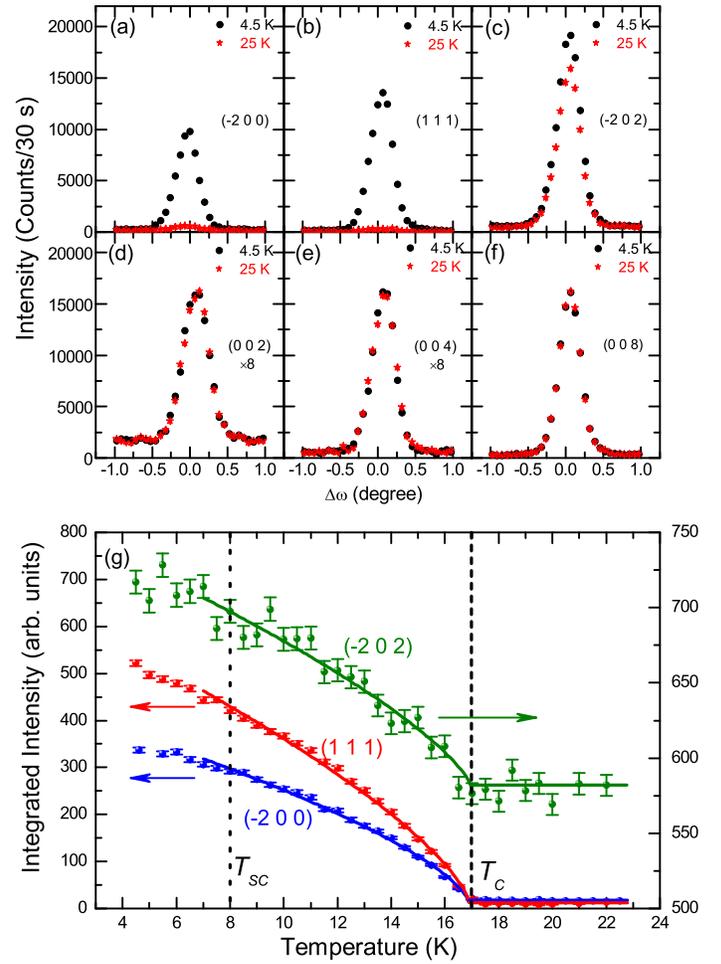}

\caption{Comparison of (a) (-2 0 0), (b) (1 1 1), (c) (-2 0 2), (d) (0 0 2), (e) (0 0 4) and (f) (0 0 8) reflections between 4.5 K and 25 K, which unambiguously indicates a ferromagnetic ordering of the Eu\textsuperscript{2+} moments along the c-direction below the magnetic ordering temperature $T_{C}$.(g) The temperature dependencies of the integrated intensities of the (-2 0 0), (1 1 1) and (-2 0 2) reflections. The solid lines represent a refinement for the temperature range between 7 K and 17 K using a power law. The ferromagnetic and superconducting transition temperatures are denoted as $T_{C}$ and $T_{SC}$ by the black vertical dashed and dotted lines, respectively.}
\end{figure}

Furthermore, two-dimensional Q-scans in both (H O L) and (H H L) planes of reciprocal space were performed at 4.5 K and are shown in Fig. 5(a) and Fig. 5(b), respectively. It is quite clear that the magnetic reflections do not appear at the (0 0 3), (-2 0 3), (1 1 0) and (1 1 2) positions corresponding to a possible antiferromagnetic ordering of the Eu\textsuperscript{2+} moments at the base temperature in the parent compound. Instead, the magnetic reflections superimpose on the nuclear peaks in both planes, indicating a magnetic propagation vector \textbf{k} = (0 0 0), consistent with the ferromagnetic ordering of the Eu\textsuperscript{2+} spins. No incommensurate reflections are observed in either plane, excluding the possibility of a helical arrangement of the Eu\textsuperscript{2+} moments in Eu(Fe\textsubscript{0.82}Co\textsubscript{0.18})\textsubscript{2}As\textsubscript{2}.

\begin{figure}
\centering{}\includegraphics{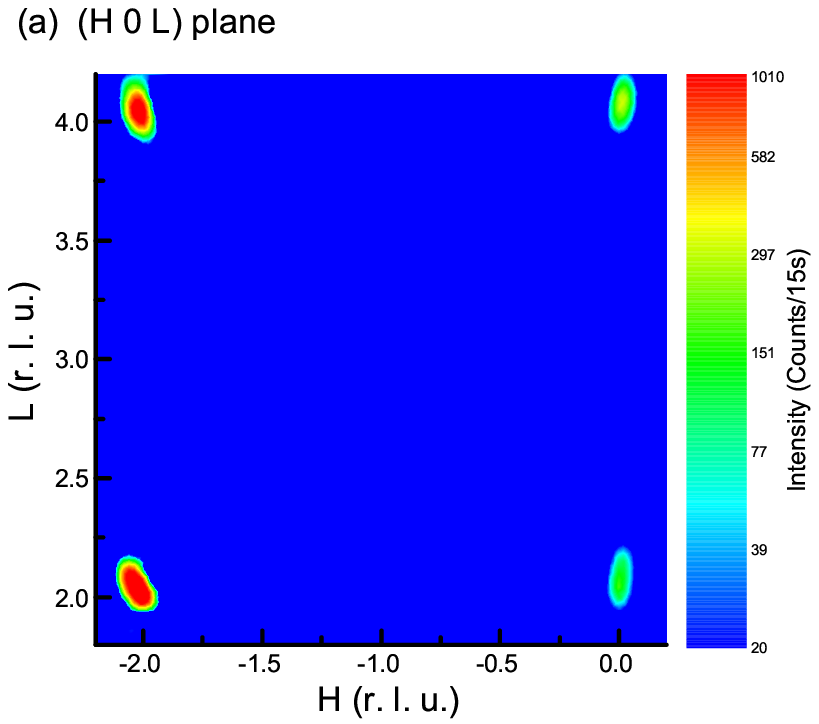}
\includegraphics{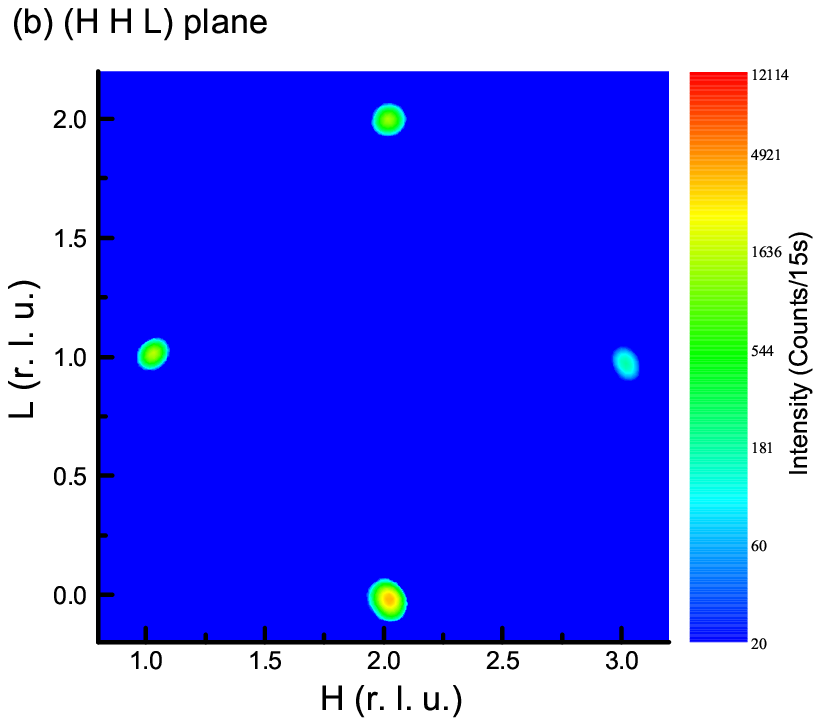}\caption{Two-dimensional Q-scans at T = 4.5 K in (a) (H 0 L) plane and (b) (H H L) plane, indicating a magnetic propagation vector \textbf{k} = (0 0 0) and excluding the possibility of both the antiferromagnetic or the helical magnetic ordering for the Eu\textsuperscript{2+} moments in Eu(Fe\textsubscript{0.82}Co\textsubscript{0.18})\textsubscript{2}As\textsubscript{2}. The intensity is shown in logarithmic scale. The magnetic reflection of the Fe moments at (-1 0 3) could not be observed here due to the short counting time.}

\end{figure}

Aside from the convincing evidences for the Eu-FM magnetic ordering, very weak magnetic reflections of the Fe\textsuperscript{2+} moments were also observed in Eu(Fe\textsubscript{0.82}Co\textsubscript{0.18})\textsubscript{2}As\textsubscript{2 }. This is quite suprising since this composition is already at a high Co concentration level as 18\%. However, the existence of a structural phase transition in this sample indicates the possibility to observe the Fe-SDW ordering since both follow each other. As shown in Fig. 6(a), at 4.5 K, a set of magnetic peaks appears at Q = (1 2 1), (1 0 3) and (1 0 5), respectively, with a propagatition vector \textbf{$\mathsf{\mathbf{k}}$} = (1 0 1), corresponding to the antiferromagnetic alignment of the Fe\textsuperscript{2+} moments along the \textit{a} axis, similar to that of the parent compound. The peak profiles of the (1 2 1) reflection at different temperatures are shown in Fig. 6(b). The peak is present below 70 K. Due to the extreme weakness of the magnetic reflections from ordering of the Fe sublattice, the temperature dependence of the peak intensity instead of the integrated intensity of the (1 2 1) reflection was measured and shown in Fig. 6(c). The peak intensity starts to increase below 70 K (denoted as SDW-ordering temperature $T_{N}$) and reaches a maximum around the superconducting transition temperature $(T_{SC}$), which is consistent with the well-established behavior of the order parameter of the Fe-SDW ordering due to its competion with SC. However, based on the limited statistics of our data, it is hard to conclude upon the presence or absence of a possible interplay between the Fe-SDW and the Eu-FM in Eu(Fe\textsubscript{0.82}Co\textsubscript{0.18})\textsubscript{2}As\textsubscript{2 }. According to the integrated intensities of three obtained Fe-magnetic reflections shown in Fig. 6(a), the size of the Fe\textsuperscript{2+} moment is roughly estimated to be \textasciitilde{} 0.15(1) $\mu_{B}$, strongly suppressed compared to 0.98(8) $\mu_{B}$ in the parent compound. \cite{Xiao_09} Thus, the Co-doping suppresses both the SDW-ordering temperature and the moment size of Fe, while the moment direction of Fe is most likely unchanged although \textit{a} and \textit{b} can't be resolved in the present case. The coexistence and competition between the Fe-SDW and the SC in Eu(Fe\textsubscript{0.82}Co\textsubscript{0.18})\textsubscript{2}As\textsubscript{2 }resembles that in Ba(Fe$_{1-x}$Co$_{x}$)\textsubscript{2}As\textsubscript{2}, \cite{pratt_09,Christianson_09} suggesting an important role of magnetism. 

\begin{figure}
\centering{}\includegraphics{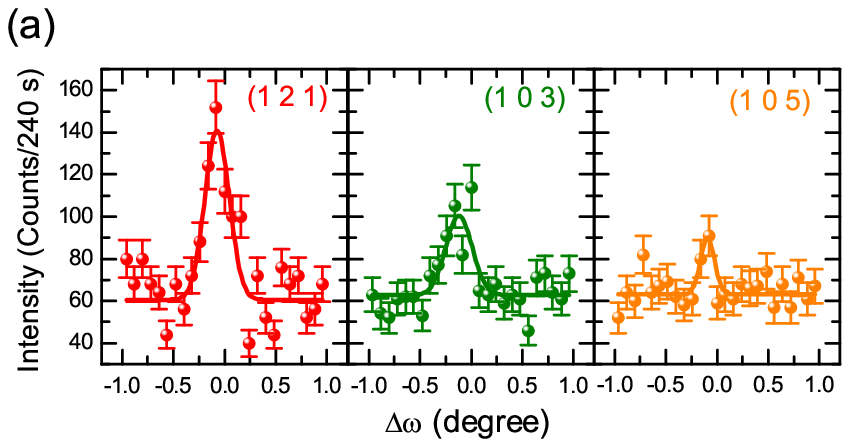}

\includegraphics{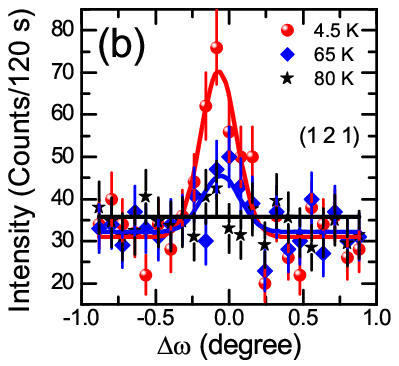}\includegraphics{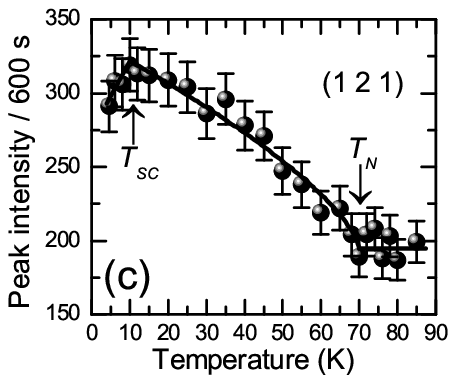}\caption{(a) The magnetic reflections at 4.5 K arsing from the SDW ordering of the Fe moments. The solid curves are refinements using Gaussian profiles. (b) The (1 2 1) reflection at 4.5 K, 65 K (just below $T_{N}$) and 80 K (above $T_{N}$), respectively. (c) The temperature dependence of the peak intenisty of the (1 2 1) reflection, in which the black solid line is a guide to the eye. }
\end{figure}

In order to precisely determine the nuclear and magnetic structures of this compound, the integrated intensities of 348 reflections at 4.5 K and 330 reflections at 25 K (both including 142 independent reflections) were collected and refined by the method of least squares after the necessary absorption correction. We use the same value for lattice constant \textbf{$a$} and \textbf{$b$} since the orthorhombic splitting is too small to be resolved. The results of the refinements are listed in Table 1.The nuclear structure in Eu(Fe\textsubscript{0.82}Co\textsubscript{0.18})\textsubscript{2}As\textsubscript{2} shows no evident difference between 4.5 K and 25 K, and the reflection set of 4.5 K could be well fitted when adding a ferromagnetic Eu\textsuperscript{2+}-moment of 6.2 $\mu_{B}$ purely along the \textit{c}-direction. This is consistent with the behavior presented above in Fig. 4 and Fig. 5. 

\begin{table}

\caption{Refined results for the nuclear and magnetic structures of Eu(Fe\textsubscript{0.82}Co\textsubscript{0.18})\textsubscript{2}As\textsubscript{2 }at 4.5 K, and also the nuclear structure at 25 K. The atomic positions are as follows: Eu, $4a$ (0, 0, 0); Fe/Co, $8f$ (0.25, 0.25, 0.25); As, $8i$(0, 0, $z$). The occupancies of Fe and Co atoms are fixed to 82\% and 18\%, respectively, according to the chemical composition determined from WDS. (Space group: $Fmmm$)}

\begin{ruledtabular} %
\begin{tabular}{cccc}
\multicolumn{2}{c}{Temperature} & 4.5 K & 25 K\tabularnewline \hline $a\,(\thickapprox b)\,(\textrm{\AA)}$ &  & 5.543(4) & 5.544(4)\tabularnewline $c\,(\textrm{\AA)}$ &  & 11.98(2)  & 12.01(2)\tabularnewline \hline Eu  & $B\,$($\buildrel_\circ \over {\mathrm{A}}$\textsuperscript{2}) & 0.28(6)  & 0.32(5) \tabularnewline  & magnetic propagation vector $\mathbf{k}$ & (0 0 0) & - \tabularnewline   & $M_{a}$($\mu_{B})$ & 0 & - \tabularnewline  & $M_{b}$($\mu_{B})$ & 0 & - \tabularnewline   & $M_{c}$($\mu_{B})$ & 6.2(1)  & - \tabularnewline Fe/Co & $B\,$($\buildrel_\circ \over {\mathrm{A}}$\textsuperscript{2}) & 0.17(4) & 0.16(3) \tabularnewline As  & $z$ & 0.362(1)  & 0.361(1) \tabularnewline  & $B\,$($\buildrel_\circ \over {\mathrm{A}}$\textsuperscript{2}) & 0.22(5)  & 0.23(4) \tabularnewline \hline  $R{}_{F^{2}}$  &  & 8.69 & 8.65 \tabularnewline $R{}_{wF^{2}}$  &  & 9.26 & 8.75\tabularnewline $R_{F}$  &  & 5.57 & 6.07\tabularnewline $\chi^{2}$ &  & 7.85 & 5.39\tabularnewline 
\end{tabular}\end{ruledtabular} 
\end{table}

The magnetic structure of Eu(Fe\textsubscript{0.82}Co\textsubscript{0.18})\textsubscript{2}As\textsubscript{2 }at 4.5 K is plotted and compared with that of the parent compound EuFe\textsubscript{2}As\textsubscript{2 }in Fig.7. The moment direction of Eu\textsuperscript{2+} flops from the\textit{ a} direction in the parent compound to the \textit{c} direction upon 18\% Co-doping into the Fe site, with the magnetic ordering pattern developing from the A-type AFM to the pure FM. This is in agreement with the spin reorientation observed by M\"{o}ssbauer spectroscopy from the \textit{a}-axis towards the \textit{c}-axis in the \textit{a-c} plane with increasing substitution of Fe by Co in Eu(Fe$_{1-x}$Co$_{x}$)\textsubscript{2}As\textsubscript{2},\cite{Blachowski_11} and also similar to the ferromagnetic ordering of Eu\textsuperscript{2+} moment along the \textit{c}-direction determined recently by resonant magnetic x-ray scattering in EuFe(As\textsubscript{0.85}P\textsubscript{0.15})\textsubscript{2 }\cite{nandi_13}. The moment size of Eu\textsuperscript{2+} in Eu(Fe\textsubscript{0.82}Co\textsubscript{0.18})\textsubscript{2}As\textsubscript{2}($\sim(6.2\pm0.1)$ $\mu_{B})$ determined here is smaller than that of the parent compound EuFe\textsubscript{2}As\textsubscript{2} ($\sim(6.7\pm0.1)$ $\mu_{B})$, suggesting the possible existence of some non-magnetic trivalent Eu\textsuperscript{3+}at a high doping level of Co, consistent with the report from the M\"{o}ssbauer study in the Eu(Fe$_{1-x}$Co$_{x}$)\textsubscript{2}As\textsubscript{2} compounds. \cite{Blachowski_11} The magnetic ordering temperature of the Eu\textsuperscript{2+} spins is slightly suppressed from $T_{N}$ = 19 K for the parent compound to $T_{C}$ = 17 K with $x$ = 0.18. The \textit{c}-direction FM ordering of Eu\textsuperscript{2+} presented
here contrasts with the canted-AFM ordering with an easy \textit{a-b} plane proposed by Tran \textit{et al.} in a single crystal with similar Co-concentration $x$ = 0.19 based on pure macroscopic measurements. \cite{tran_12} The microspopic bulk probe of neutron diffraction
in our study, nevertheless, provides convincing evidence for the ferromagnetic ordering of the Eu\textsuperscript{2+} moments. Regarding the magnetic structure of the Fe sublattice, the moment size of Fe is significantly suppressed, from 0.98(8) $\mu_{B}$ in the parent compound to \textasciitilde{} 0.15(1) $\mu_{B}$ with $x$ = 0.18, while probably keeping the moment direction of Fe unchanged with the Co-doping. 

\begin{figure}
\centering{}\includegraphics[scale=0.36]{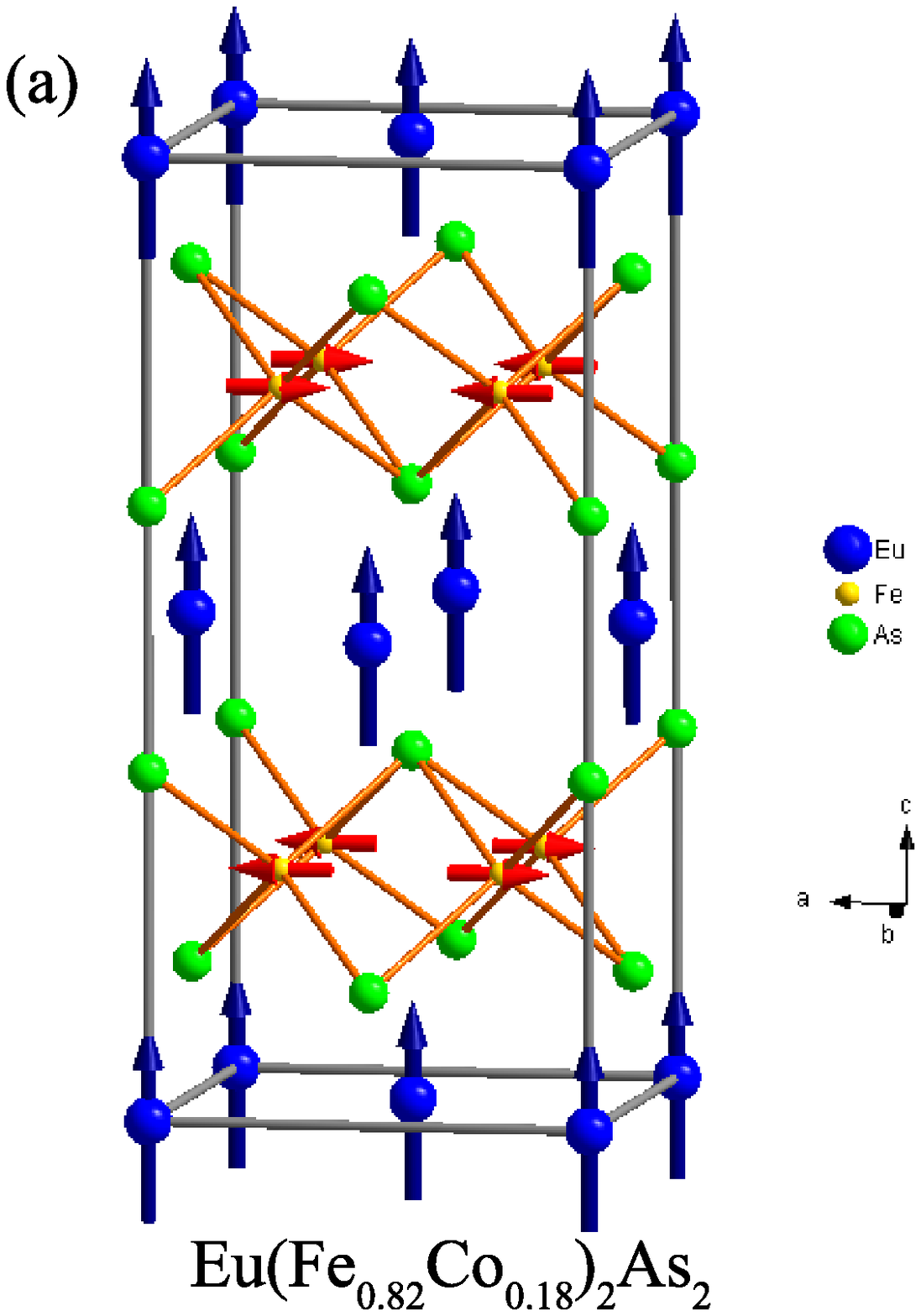}\includegraphics[scale=0.36]{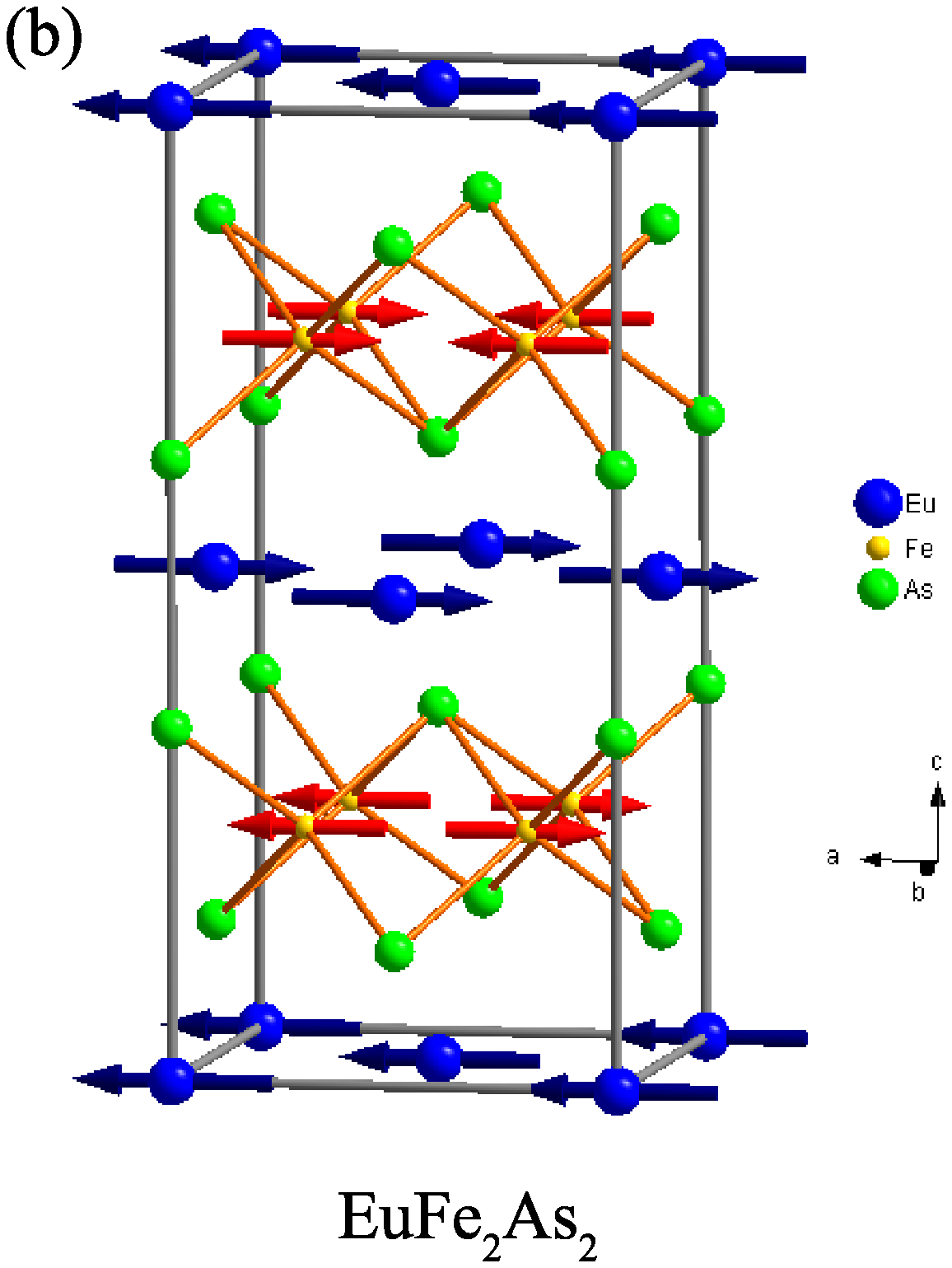}

\caption{The magnetic structure of (a) Eu(Fe\textsubscript{0.82}Co\textsubscript{0.18})\textsubscript{2}As\textsubscript{2 }and (b) the parent compound EuFe\textsubscript{2}As\textsubscript{2 }at base temperature. }
\end{figure}

Based on all the results above, we come to two important conclusions. First, the Eu\textsuperscript{2+} moments exhibit a long-range ferromagnetic ordering in the superconducting Eu(Fe\textsubscript{0.82}Co\textsubscript{0.18})\textsubscript{2}As\textsubscript{2 }crystal.
This proves to be a common feature for a number of doped EuFe\textsubscript{2}As\textsubscript{2} compounds with superconductivity, for instance, with P-doping, \cite{ren_EuP_09,nandi_13} Ru-doping, \cite{jiao_EPL09,jiao_EuRu12} and Co-doping presented here. Due to the small superconducting volume in the crystal and the dominance of the Eu-FM, it is difficult to conclude about the interplay between the SC and the Eu-FM in Eu(Fe\textsubscript{0.82}Co\textsubscript{0.18})\textsubscript{2}As\textsubscript{2 }. \cite{jiao_EPL09} Second, the Fe-SDW and the SC coexist and compete with each other in Eu(Fe\textsubscript{0.82}Co\textsubscript{0.18})\textsubscript{2}As\textsubscript{2 }. Although it is already at a high doping level  with a $T_{SC}$ = 8 K, the antiferromagnetism from the Fe-SDW as well as the structural phase transtition still survive. Both the structural phase transition ($T_{S}$ = 90 K) and the Fe-SDW phase transition ($T_{N}$ = 70 K) are significantly suppressed compared to the parent compound EuFe\textsubscript{2}As\textsubscript{2}, but splitted by 20 K, similar to the observation in Ba(Fe$_{1-x}$Co$_{x}$)\textsubscript{2}As\textsubscript{2}.\cite{pratt_09,Christianson_09} This interplay between these two order parameters is already well studied in other ``122'' families and attributed to the competition for the shared electronic denisty of states common to both Fermi surface gaps caused by the Fe-SDW and the SC. Moreover, the critical point at which the Fe-SDW is completely suppressed in Eu(Fe$_{1-x}$Co$_{x}$)\textsubscript{2}As\textsubscript{2} (possibly larger than $x$ = 0.2) is considerable higher than that in Ba(Fe$_{1-x}$Co$_{x}$)\textsubscript{2}As\textsubscript{2} ($x$ $\thickapprox$ 0.065) \cite{pratt_09}, indicating a considerable influence of the rare-earth element Eu on the magnetism of Fe sublattice.

\section{Conclusion}

In summary, the magnetic structure of superconducting Eu(Fe\textsubscript{0.82}Co\textsubscript{0.18})\textsubscript{2}As\textsubscript{2 }is unambiguously determined by single crystal neutron diffraction. A long-range ferromagnetic order of the Eu\textsuperscript{2+} moments along the c-axis is revealed below the magnetic transition temperature $T_{C}$ = 17 K. No incommensurate magnetic reflections corresponding to the helical arrangement of the Eu\textsuperscript{2+} spins is observed for this crystal. In addition, the antiferromagnetism of the Fe\textsuperscript{2+ }moments still survives as does the tetragonal-to-orthorhombic structural phase transition, although the transition temperatures of the Fe-spin density wave (SDW) order and the structural phase transition are significantly suppressed to $T_{N}$ = 70 K and $T_{S}$ = 90 K, respectively, compared to the parent compound EuFe\textsubscript{2}As\textsubscript{2}. We present the microscopic evidence for the coexistence of the Eu-FM and the Fe-SDW in the superconducting crystal, which is quite rare and unusual. The SC competes with the Fe-SDW in Eu(Fe\textsubscript{0.82}Co\textsubscript{0.18})\textsubscript{2}As\textsubscript{2 }, similar to the behavior found in the Ba(Fe$_{1-x}$Co$_{x}$)\textsubscript{2}As\textsubscript{2}. However, due to the small superconducting volume in the crystal and the dominance of the Eu-FM, it is difficult to conclude about the interplay between the SC and the Eu-FM in Eu(Fe\textsubscript{0.82}Co\textsubscript{0.18})\textsubscript{2}As\textsubscript{2. }Moreover, the critical point at which the Fe-SDW is completely suppressed in Eu(Fe$_{1-x}$Co$_{x}$)\textsubscript{2}As\textsubscript{2} is considerable higher than in Ba(Fe$_{1-x}$Co$_{x}$)\textsubscript{2}As\textsubscript{2}, indicating a considerable influence of the rare-earth element Eu on the magnetism of Fe sublattice.

\begin{acknowledgments}
This work is based on experiments performed at the Swiss spallation neutron source SINQ, Paul Scherrer Institute, Villigen, Switzerland. W. T. Jin would like to thank J. Schefer for the help at the TriCS beamline, and B. Schmitz and J. Persson for their technical assistance. Z. B. acknowledges the financial support from the National Science Center of Poland, Grant 2011/01/B/ST5/06397.
\end{acknowledgments}

\end{document}